\documentclass[a4paper,11pt]{article}
\usepackage{pos}
\usepackage{hyperref}

\newcommand{\pbp}{\langle \bar{\psi} \psi \rangle}
\newcommand{\be}{\begin{equation}}
\newcommand{\ee}{\end{equation}}
\newcommand{\mygraph}[1]{\includegraphics[width=#1,keepaspectratio]}
\newcommand{\msub}[1]{\ensuremath _{\mbox{\tiny #1}}}
\newcommand{\nii}{n_i}
\newcommand{\naa}{n_a}

\title{$U(1)_A$ Breaking in Hot QCD in the Chiral Limit}
\author*[a,b]{Tamas G.\ Kovacs}
\affiliation[a]{Institute of Physics and Astronomy,
                ELTE E\"otv\"os Lor\'and University\\
                P\'azm\'any P\'eter s\'et\'any 1/A, 
                H-1117 Budapest, Hungary
}

\affiliation[b]{Institute for Nuclear Research  
                Bem t\'er 18/c, H-4026 Debrecen, Hungary
}

\emailAdd{tamas.gyorgy.kovacs@ttk.elte.hu}
\abstract{We propose a simple instanton-based random matrix model of hot QCD
  that in the quenched case precisely reproduces the distribution of the
  lowest lattice overlap Dirac eigenvalues. Even after including dynamical
  quarks the model can be easily simulated in volumes and for quark masses
  that will be out of reach for direct lattice simulations in the foreseeable
  future. Our simulations show that quantities connected to the $U(1)_A$ and
  $SU(N_f)_A$ chiral symmetry are dominated by eigenvalues in a peak of the
  spectral density that becomes singular at zero in the thermodynamic
  limit. This spectral peak turns out to be produced by an ideal instanton
  gas. By generalizing Banks-Casher type integrals for the singular spectral
  density, definite predictions can be given for physical quantities that are
  essential to test chiral symmetry breaking, but presently impossible to
  compute reliably with direct lattice simulations.}

\FullConference{The 41st International Symposium on Lattice Field Theory
  (LATTICE2024)\\ 
 28 July - 3 August 2024\\
Liverpool, UK\\}

\begin{document}
\maketitle

\section{Introduction}

Quantum chromodynamics (QCD), the theory of strong interactions has an
approximate $SU(2)_L \times SU(2)_R \times U(1)_V \times U(1)_A$ symmetry. The
flavor non-singlet vector symmetry is a result of the approximate equality of
the masses of the two light quarks, the $u$ and $d$, and the axial symmetries
are there because both $m_u$ and $m_d$ are much lighter than the relevant QCD
scale. Classically, the symmetries would all be exact if $m_u=m_d=0$ were to
hold exactly.

\begin{figure}
  \mygraph{0.45\textwidth}{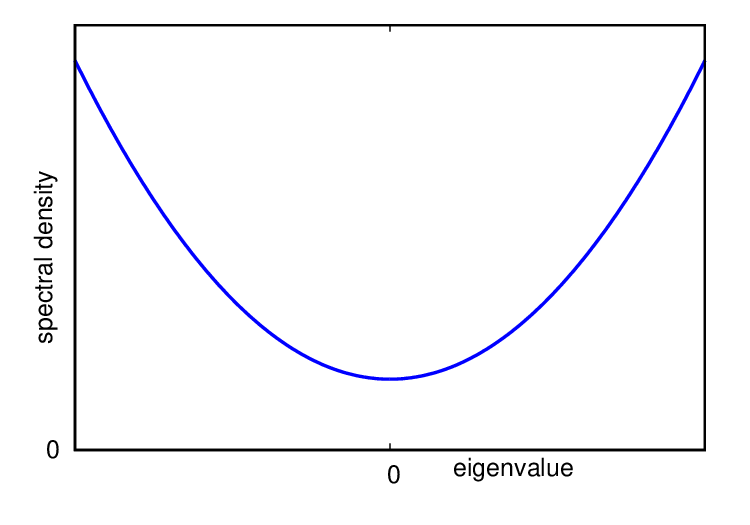} \hfill 
  \mygraph{0.45\textwidth}{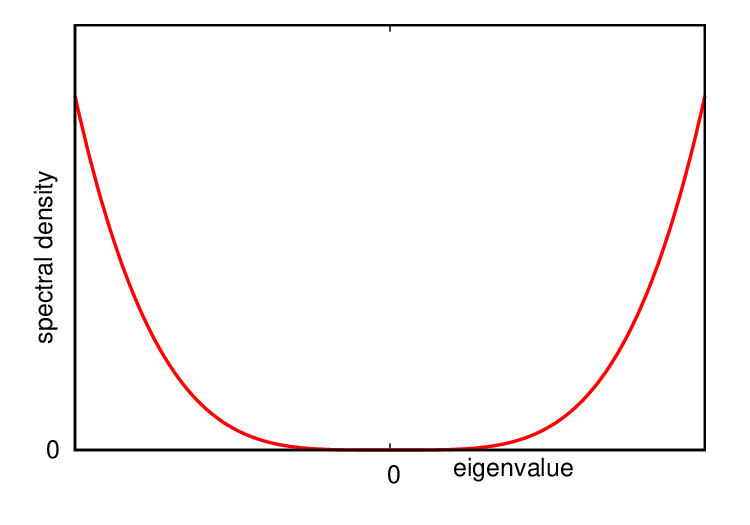}
  \caption{ \label{fig:spdenshlT} A schematic representation of the spectral
    density of the Dirac operator below the critical temperature in the
    hadronic phase (left), and above the critical temperature, in the
    quark-gluon plasma phase (right).}
\end{figure}

At low temperature the $SU(2)_L \times SU(2)_R$ symmetry is broken
spontaneously, and it is only the diagonal vector subgroup of the symmetry
that remains intact. However, at finite temperature, the transition to the
quark-gluon plasma phase results in the restoration of the axial part of the
symmetry $SU(2)_A$. The order parameter characterizing this transition is the
chiral condensate $\pbp$. This can also be written in terms of the spectrum of
the Dirac operator as
\be
  \pbp = \;\; \propto \;\; 
      \frac{1}{V} \sum_k \frac{1}{i \lambda_k +m} \;\; \propto \;\; 
      \int_{-\Lambda}^{\Lambda} d\lambda
      \frac{m }{\lambda^2+m^2} \, \rho(\lambda)  \;\;
      \xrightarrow[m\rightarrow 0]{} \;\; \rho(0),
      \label{eq:bankscasher}
\ee
where $i \lambda_k$ are the eigenvalues of the Dirac operator, and the sum over
the eigenvalues can be written in terms of the Dirac spectral density
$\rho(\lambda)$, using the symmetry $i\lambda \leftrightarrow -i\lambda$ of
the spectrum. Due to the $\frac{m}{\lambda^2+m^2}$ factor, for small quark
masses most of the contribution to the integral comes from the spectral region
$|\lambda| \lesssim m$. As a result, in the chiral limit the only contribution
is given by the spectral density at zero, which is the Banks-Casher formula
\cite{Banks:1979yr}. In this way, the spectral density at zero can also serve
as an oder parameter of the transition. In Fig.~\ref{fig:spdenshlT} we show a
schematic plot of the spectral density in the low temperature, symmetry broken
phase and in the high temperature phase where the symmetry is restored.

The spectrum of the Dirac operator can also be calculated in lattice
simulations, and for a long time lattice results were in accordance with
the above expectations; in the high temperature phase the spectral density
appeared to vanish at zero virtuality. This was the situation until the
appearance of chirally symmetric lattice Dirac operators, in particular the
overlap \cite{Narayanan:1994gw}. The first hint at a more complicated picture
appeared in Ref.~\cite{Edwards:1999zm}, where a peculiar spike was found
at high temperature in the spectrum at zero of the chirally symmetric overlap
Dirac operator. This was a quenched simulation on rather coarse lattices, and
for a long time this spectral spike was largely ignored, most likely because
it was considered a quenched or coarse lattice artifact.

Recently more detailed studies of the spectral spike appeared. In particular,
evidence was found that it is neither a quenched nor a coarse lattice
artifact, as the spike was found to be present on finer lattices both with and
without dynamical quarks \cite{Alexandru:2015fxa}. It was also suggested that
the spectral spike was singular in the thermodynamic limit, and could
signal a separate phase of QCD, intermediate between the finite temperature
crossover and the even higher temperature perturbative regime
\cite{Alexandru:2019gdm}. The presence of the spike was also independently
verified \cite{Kaczmarek:2021ser,Ding:2020xlj}, and shown also for sea quarks
lighter than the physical $u$ and $d$ quarks \cite{Kaczmarek:2023bxb}. More
recently in a large scale study with staggered sea and valence quarks, the
spectral peak was seen to be present in the continuum limit
\cite{Alexandru:2024tel}. Quark eigenstates in the peak were found to exhibit
nontrivial localization properties, hinting at the possible presence of
another mobility edge, very close to zero in the spectrum
\cite{Alexandru:2021pap,Alexandru:2021xoi,Meng:2023nxf}, in addition to the
already well established mobility edge higher up in the spectrum, separating
localized modes from the bulk \cite{Giordano:2021qav}.

All the works cited in the above paragraph used staggered or other non-chiral
sea quarks, although some utilized chiral (overlap) quarks for the
valence. Since the spectral peak is a narrow structure close to zero, it is
important for the Dirac operator to properly resolve the small Dirac
eigenvalues, for which exact chiral symmetry is needed. This point is
particularly important, since after integrating out the quarks, the partition
function of QCD reads
\be
Z=\int {\cal\!D}U \;\;
      \prod_f \det\left( D[U] + m\msub{f} \right)
      \cdot \mbox{e}^{-S\msub{g}[U] },
\ee
where $U$ is the gauge field configuration, $D[U]$ is the covariant Dirac
operator and $S\msub{g}$ is the gauge action. If the quark mass is small, the
quark determinant is expected to suppress configurations with many small Dirac
eigenvalues which might result in the disappearance of the spectral peak at
zero. There are also studies using (close to) chiral quarks, namely domain
wall quarks. Indeed, their conclusion is that for small but finite quark
masses, the spectral peak is so suppressed that it cannot be detected at all
\cite{Cossu:2013uua,Tomiya:2016jwr,Aoki:2020noz}. For the most recent updates
see \cite{JLQCD:2024xey,Ward:2024wze}.

The spectral peak is not only important for the Banks-Casher relation, and the
restoration of the flavor non-singlet chiral symmetry. As discussed in many of
the above cited works (see also \cite{Azcoiti:2023xvu}), it also determines
the fate of the $U(1)_A$ axial symmetry. It can be shown that if in the chiral
limit the Dirac spectral density develops a gap at zero, or even if it is
analytic at zero, the effects of the anomalous $U(1)_A$ symmetry breaking
cannot be seen in scalar and pseudoscalar meson correlators
\cite{Aoki:2012yj}. The fate of the spectral spike is thus also essential for
the $U(1)_A$ symmetry.

In the present paper I would like to ask and answer the following three
questions. What is the physical origin of the spectral peak at zero that
appears in high temperature QCD? How is it suppressed by the quark determinant
if light dynamical quarks are present? How does the peak influence the fate of
the flavor singlet and the flavor non-singlet axial symmetries in the chiral
limit?

\section{The spectral peak in quenched QCD}

\begin{figure}
  \begin{centering}
  \mygraph{0.75\textwidth}{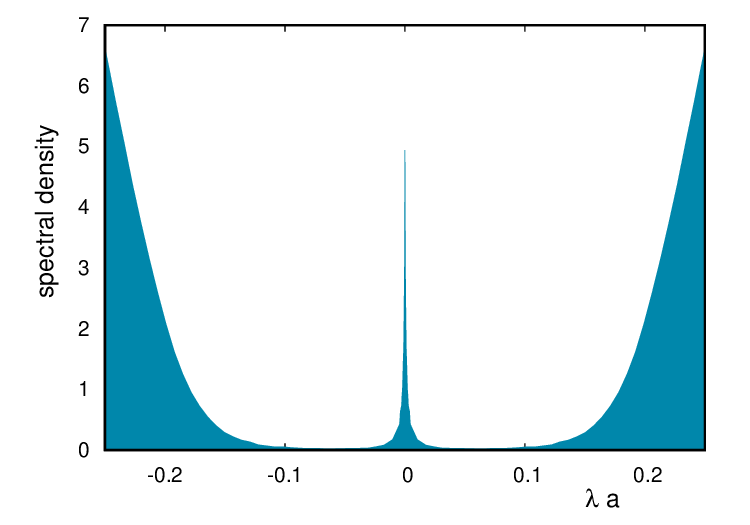}  
  \caption{ \label{fig:qspec} The spectral density of the overlap Dirac
    operator on a set of quenched gauge configurations with temporal size
    $N_t=8$ and temperature $T=1.05T_c$. The exact zero eigenvalues have been
    removed, they would show up as a delta function at zero. }
\end{centering}
\end{figure}

Since the quenched case is much simpler and there is general consensus about
the presence of the spectral spike there, we start the discussion with the
quenched theory. In Fig.~\ref{fig:qspec} we show a typical example of the
overlap Dirac spectrum on quenched gauge field configurations slightly above
the critical temperature. In the quenched case there is a genuine phase
transition and thus $T_c$ is well defined. The quenched simulation in question
was done with the Wilson action at $\beta=6.13$, corresponding to $T=1.05T_c$
for the temporal lattice size $N_t=8$. Since we are using the overlap Dirac
operator, there are also exact zero eigenvalues in the spectrum, but those
have been removed, they are not shown in the figure.

Starting from larger eigenvalues and going toward zero, the spectral density
falls sharply, and it appears to go to zero, as expected in the symmetry
restored phase. This behavior can also be qualitatively understood by noting
that at finite temperature, due to the antiperiodic temporal boundary
condition, the spectrum of free fermions have a gap around zero, equal to the
lowest Matsubara frequency. If the fermions interact with the gauge field, the
question is how the correlation length compares to the temporal box size,
i.e.\ the inverse temperature. If the correlation length is smaller than
1$/T$, then the fermions effectively ``do not feel'' the boundary condition,
and there is no trace of the Matsubara gap; the spectrum extends with a
nonzero density all the way down to the origin. In QCD this is exactly what
happens in the chirally broken phase. On the other hand, if the correlation
length is larger than $1/T$, even though there is no sharp Matsubara gap, the
part of the spectrum below the lowest Matsubara frequency becomes heavily
depleted.

This is exactly what we see in Fig.~\ref{fig:qspec}. However, at very small
eigenvalues the spectral density starts to increase again, forming a sharp
peak of near zero eigenvalues. In view of the foregoing discussion, this
unnatural proliferation of small eigenvalues is certainly unexpected, and
calls for an explanation. It was already suggested in
Ref.~\cite{Narayanan:1994gw} that the presence of instantons and
anti-instantons might be responsible for the excess of small Dirac
eigenvalues. Indeed, due to the Atiyah-Singer index theorem, if the gauge
field configuration has topological charge $Q$, then the Dirac operator has
(at least) $|Q|$ exact zero eigenvalues with chirality plus or minus one,
depending on the sign of $Q$ \cite{Atiyah:1971rm}. In particular, each
instanton and anti-instanton carries a zero mode of given chirality. In the
presence of $\nii$ instantons and $\naa$ anti-instantons there are
$|\nii - \naa |$ exact zero modes, while the rest of the would be zero modes
will mix and split around zero. The splitting is governed by the distance
between the topological objects and also their relative orientation in group
space \cite{Schafer:1996wv}.

An important quantity characterizing the fluctuations of the topological
charge is the topological susceptibility defined as
\be
  \chi = \frac{1}{V} \langle Q^2 \rangle,
\ee
where $V$ is the four-volume of the system and the angled brackets denote
expectation with respect to the path integral. Above the finite temperature
transition the topological susceptibility falls sharply, and instantons form a
dilute gas. The typical separation among instantons and anti-instantons is
thus large, resulting in a small splitting of near zero modes around
zero. This mechanism is a promising candidate to explain the spike in the
spectral density.

At this point we have to make a remark about our use of the terminology. By
{\em instanton} one typically means an exact solution of the Euclidean
equation of motion, carrying and integer topological charge. There is ample
evidence that in gauge configurations dominating the path integral,
topological charge does not come in the form of pristine instantons (or
calorons, the finite temperature exact solutions), not even well above the
critical temperature. This can be seen from the fact that the dilute instanton
gas approximation that assumes small fluctuations around the exact solution,
cannot describe the dependence of the topological susceptibility on the
temperature, seen in lattice simulations
\cite{Borsanyi:2016ksw,Bonati:2015vqz,Petreczky:2016vrs}. See also
Ref.\cite{Boccaletti:2020mxu} for a recent update on the perturbative
instanton calculation. In spite of all this, for simplicity we will use the
word {\em instanton}, but by this we mean only well defined lumps of unit
topological charge. We will show that the spike in the Dirac spectrum above
$T_c$ can be understood by assuming that the topological charge comes in the
form of well separated lumps of unit charge.

\section{Random matrix model of the zero mode zone of the Dirac operator}

In what follows we would like to build a simple model of the subspace of the
Dirac operator spanned by the would be zero modes that we call the zero mode
zone (ZMZ). Our starting point is the simplest nontrivial configuration, that
of an instanton and an anti-instanton at a distance $r$ apart. In this case
the ZMZ can be represented by the $2 \times 2$ matrix
\be
 \hspace{-15ex}  D(A)\msub{zmz} \; = \; \left(
    \begin{array}{cc}
       0  &  iw \\
      iw  &   0
    \end{array}
   \right), % \hspace{10ex}   w \; \propto \; \mbox{e}^{-\pi T r}
\ee
where
\be
w \; \propto \; \mbox{e}^{-\pi T r}
\ee
is the mixing between the instanton and anti-instanton zero mode. Since at
high temperature the zero modes are exponentially localized with a
localization length of $1/\pi T$ \cite{Gross:1980br,GarciaPerez:1999ux}, the
mixing is also expected to be exponentially small in the distance $r$.

This can be easily generalized for the case when there are $\nii$ instantons
and $\naa$ instantons present. Then the matrix of the Dirac operator in the
ZMZ has the form

       \hspace{44ex} $\nii$ \hspace{9ex} $\naa$ \\
         \hspace*{44ex} $\overbrace{\hspace{9ex}} 
          \overbrace{\hspace{13ex}}$ \\[-2ex]
\be
              \hspace{14ex} D(A)\msub{zmz} =  
          \left(
             \begin{array}{ccc|ccccc}
               &   &             & & &    & & \\
               & 0 &             & & & iW & & \\
               &   &             & & &    & & \\
               \hline
               &   &             & & &    & & \\
               &   &             & & &    & & \\
               & iW^\dagger &    & & &  0 & & \\
               &   &             & & &   & & \\
               &   &             & & &   & & \\
             \end{array}
           \right)
\ee
with two diagonal blocks of zeros and with the matrix elements
\be
  w_{ij} = A \times \exp(-\pi T \cdot r_{ij}),
  \label{eq:expoverlap}
\ee
$r_{ij}$ denoting the distance of instanton $i$ and anti-instanton
$j$. The rank of this matrix is easily seen to be such that it has
$|\nii-\naa|$ exact zero eigenvalues and the magnitude of the rest of the
eigenvalues is controlled by the exponentially small off-diagonal matrix
elements $w_{ij}$. The matrix also has exact chiral symmetry, as it
anticommutes with $\gamma_5$ which in this basis takes a diagonal form with
$\nii$ $-1$-s and $\naa$ $+1$-s in the diagonal. Given any instanton
configuration, we can construct the matrix using the locations of the
instantons and anti-instantons.

To complete the model we still have to decide how to choose the number of
(anti-)instantons and their locations. It has been established that in
quenched QCD, above $T_c$ the distribution of the topological charge
\cite{Bonati:2013tt}, as well as that of the number of eigenvalues in the
spike are consistent with a noninteracting, free instanton gas
\cite{Vig:2021oyt}. This suggests that we can choose the locations of the
topological objects independently with a uniform spatial distribution. Since
above $T_c$ the typical instanton size is expected to be comparable to the
temporal box size, we adopt a dimensionally reduced picture by choosing the
instanton locations and measuring their distances in a 3D box of spatial size
$L^3$, completely ignoring the temporal dimension. Finally, in a free
instanton gas the distribution of the number of instantons and anti-instantons
follows independent and identical Poisson distributions. The probability of
having $\nii$ instantons and $\naa$ anti-instantons is thus
\be
     p(\nii,\naa) = \mbox{e}^{-\chi V} \times
      \frac{(\chi V/2)^{\nii}}{\nii !} \times
      \frac{(\chi V/2)^{\naa}}{\naa !},
\ee
where $\chi$ is the topological susceptibility, $V$ is the four-volume. This
completes the definition of our model. If the two parameters, the topological
susceptibility and the prefactor $A$ in Eq.~(\ref{eq:expoverlap}) is known, we
can produce an ensemble of random matrices that model the zero mode zone of
the lattice Dirac operator.

\begin{figure}
  \begin{centering}
  \mygraph{0.75\textwidth}{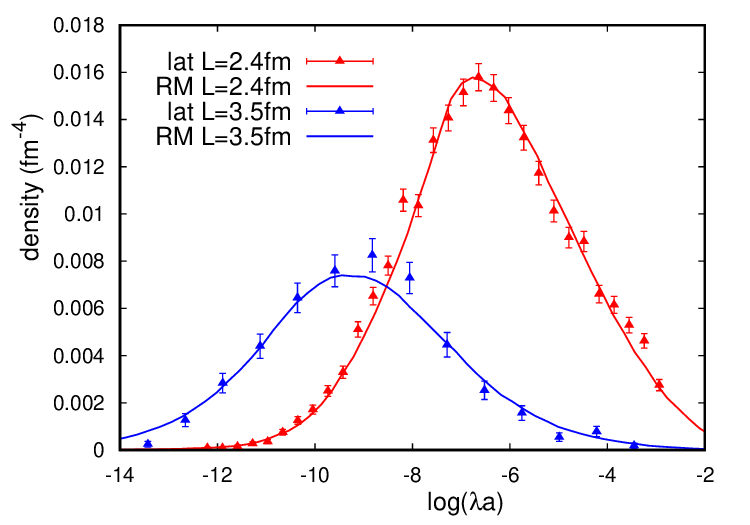} \hfill 
  \caption{\label{fig:fitA} The distribution of the lowest overlap Dirac
    eigenvalue on two ensembles of quenched gauge configurations with
    different volumes. The temperature is $T=1.1T_c$ in both cases, and for a
    better resolution of the small eigenvalues we plotted the distribution of
    the (natural) log of the eigenvalues. The triangles represent the lattice
    data, the continuous lines are calculated from the random matrix model. We
    used the smaller volume for fitting the parameter $A$, the larger volume
    is a prediction of the model without any further fitting.}
 \end{centering}
\end{figure}

To determine the two parameters, we computed the lowest part of the Dirac
spectrum on an ensemble of $32^3 \times 8$ quenched lattice configurations at
$T=1.1T_c$. We determined the topological susceptibility by counting the
number of exact zero eigenvalues, giving the topological charge on each
configuration. The parameter $A$ can be fitted to any feature of the lattice
overlap Dirac spectrum. We chose to look at the distribution of the lowest
nonzero eigenvalues, and could obtain a good fit of the whole distribution
with the choice $A=0.35$. This completes the fixing of the parameters, and now
we can test how well the resulting model describes the spectrum in different
situations. A possible test is the distribution of the lowest eigenvalue on a
larger volume. In Fig.~\ref{fig:fitA} we compare the distribution of the
lowest eigenvalue in the random matrix model to that of the lattice Dirac
operator for two different volumes. The smaller volume with linear size
$L=2.5$~fm was used for fitting $A$, the larger volume, $L=3.5$~fm is a
prediction of the model that appears to agree with the lattice
distribution. Several other tests can be performed for the distribution of
higher eigenvalues in different volumes, also with restriction to a given
topological charge sector, and in all cases the prediction of the model agrees
with the lattice results. This indicates that the random matrix model, based on
the free instanton picture gives an excellent description of the zero mode
zone of the lattice overlap spectrum in the quenched case. 

We would like to remark that the instanton-based random matrix model we
propose here is not new. Already a long time ago it was extensively used in
the context of instanton liquid models (see e.g.\
\cite{Schafer:1996wv,Shuryak:1992pi}). What is new here is that our model is
much simpler than the previous ones, as we use a dimensionally reduced 3D
picture, ignore gauge interactions among topological lumps, and completely
neglect the orientation of instantons in group space, as well as their
nontrivial size distribution. These are all reasonable assumptions at high
temperature in the quenched case. The other novelty of our model is that after
fitting its two parameters, it can precisely describe the detailed features of
the lattice overlap Dirac spectrum.

\section{Full QCD, including dynamical quarks}

So far we presented a simple random matrix model for describing the zero mode
zone of the lattice overlap Dirac operator in the quenched case. Our goal in
this section is to include dynamical quarks in the model. On the lattice this
is done by including the fermion determinant $\det(D+m)^{N\msub{f}}$ in the
Boltzmann weight of each configuration. For simplicity, we will assume $N_f$
degenerate quark flavors with mass $m$, but the generalization to different
masses is straightforward. If the quark determinant is written in terms of the
Dirac eigenvalues, it can be split into a product of two contributions, one
coming from the eigenvalues in the zero mode zone, and the contribution of the
rest of the spectrum that we call the bulk as
\be 
  \det(D+m) = \prod\msub{zmz} (\lambda_i + m) \times
    \prod\msub{bulk} (\lambda_i + m). 
\ee
The main observation here is that the bulk is separated from the ZMZ by a
strongly depleted region in the spectrum, and the contribution of the bulk is
not expected to be correlated with that of the ZMZ. Therefore, when computing
expectations pertaining to the ZMZ, the bulk contribution gives only a
constant factor, canceling in the expectations. In this way, when computing
physical quantities dominated by the ZMZ, the contribution of the bulk to the
determinant can be omitted. On the other hand, the contribution to the
determinant of the ZMZ can be consistently computed within our random matrix
model, and it can be added as an additional weight for each quenched instanton
configuration. So the random matrix model of the ZMZ of full QCD will have the
weight
\be
  P(\nii, \naa) \propto \underbrace{\mbox{e}^{-\chi\msub{0}V}
      \frac{1}{\nii !} \frac{1}{\naa !}
      \left( \frac{\chi\msub{0} V}{2}\right)^{\nii + \naa}}_{\mbox{\scriptsize
        free instanton gas with random locations}}
    \times \det(D+m)^{N\msub{f}}
  \label{eq:iaprob}
\ee
for a given instanton configuration with $\nii$ instantons and $\naa$
anti-instantons. Here $\chi_0$ is the quenched susceptibility, and the first
part of the weight comes from the Poisson distributions of the free instanton
gas, and the determinant of the random matrices is the extra reweighting
factor due to the sea quarks. Although we have not indicated it explicitly,
the determinant depends on the location of the topological objects.

\section{Simulation of the random matrix model of the ZMZ of full QCD}

\begin{figure}
  \begin{centering}
  \mygraph{0.75\textwidth}{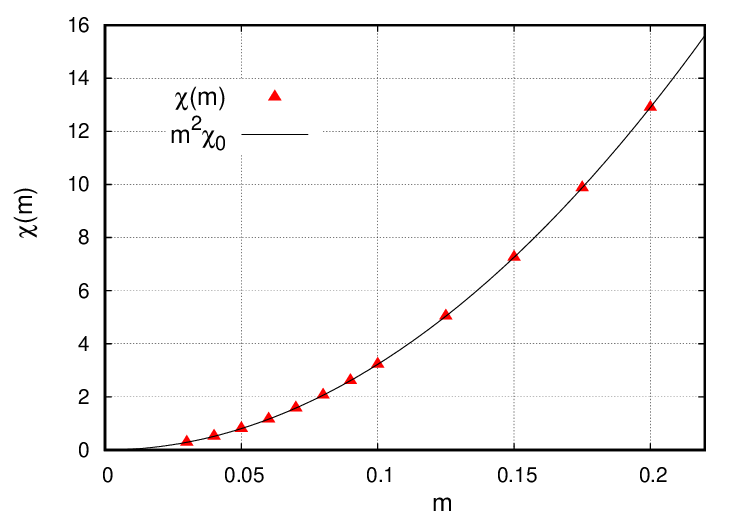} \hfill 
  \caption{\label{fig:simchi}  The topological susceptibility as a function of
  the quark mass, obtained from a simulation of the random matrix model,
  including the reweighting with the quark determinant for two degenerate
  quark flavors. The simulation data is plotted with triangles, and the
  continuous curve is not a fit, but the quadratic function $m^2 \chi_0$,
  where $\chi_0$ is the quenched topological susceptibility, the parameter of
  the model. Without the reweighting with the determinant, this would be the
  topological susceptibility.}
 \end{centering}
\end{figure}

In this section we present the results of a simulation of the random matrix
model for two degenerate flavors of dynamical quarks. In Fig.~\ref{fig:simchi}
we show the topological susceptibility as a function of the quark mass. The
triangles are the simulation data, and the continuous line is not a fit, but
the quadratic function $\chi_0 m^2$, and it describes the simulation data
perfectly.

To understand why this happens, let us first assume that the quark mass is
much larger than the typical eigenvalues in the ZMZ. In this case the quark
determinant can be approximated as
\be
  \prod_{\nii,\naa} (\lambda_i +m)^{N_f} \;\; \approx \;\;
   m^{N_f (\nii + \naa )},
\ee
which means that in this approximation the determinant does not depend on the
instanton locations, only on their numbers. This, in turn, implies that the
reweighting with the determinant does not introduce any interaction among the
topological objects. Indeed, substituting the approximated determinant into
Eq.~(\ref{eq:iaprob}), the $m^{N_f (\nii + \naa )}$ factor can be absorbed into
the Poisson distributions as
\be
  P(\nii,\naa) \;\; \propto \;\; \left( \frac{\chi\msub{0}
      V}{2}\right)^{\nii + \naa} 
    \!\! \times \det(D+m)^{N\msub{f}} \;\;\; \approx \;\;\; 
    \left( \frac{m^{N\msub{f}} \chi\msub{0} V}{2}\right)^{\nii + \naa}
    \label{eq:approxdet1}
\ee
which, after proper normalization, results again in Poisson distributions for
the number of instantons and anti-instantons, but with a density suppressed by
the quark mass as $\chi\msub{0} \rightarrow m^{N\msub{f}}\chi\msub{0}$. In
other words, if $|\lambda_i| \ll m$, then the matrix model with dynamical quarks
still describes a free instanton gas, but with a smaller topological
susceptibility.

What happens if the quark mass becomes smaller? To understand this, we write
the determinant as
\be
  \det(D+m)^{N\msub{f}} \;\;\; =  \;\;\; m^{N_f (\nii + \naa )} \times
  \prod_i \left( 1 +  \frac{|\lambda_i|^2}{m^2} \right),
  \label{eq:approxdet2}
\ee
where the last product is over the nonzero complex conjugate pairs of
eigenvalues of the random matrix, representing the Dirac operator. If the
quark mass is small, the power of the quark mass heavily suppresses those
configurations that have many instantons. The configurations contributing to
expectations have only a small number of instantons that are far apart. Since
the matrix elements of the random matrix are exponentially small in the
distance between the topological objects, the typical eigenvalues also become
smaller and smaller as the instanton gas becomes more dilute. Consequently, no
matter how small the quark mass is, the eigenvalues of the random matrices
always remain much smaller than the quark mass, the last product of
Eq.~(\ref{eq:approxdet2}) is essentially equal to unity, and the approximation
in Eq.~(\ref{eq:approxdet1}) remains valid even in the chiral limit. This
explains our finding that for two quark flavors the topological susceptibility
is proportional to $m^2$, and the lowest part of the Dirac spectrum can be
understood as the zero mode zone of a free instanton gas.

\section{The nature of the spectral spike}

\begin{figure}
  \begin{centering}
  \mygraph{0.75\textwidth}{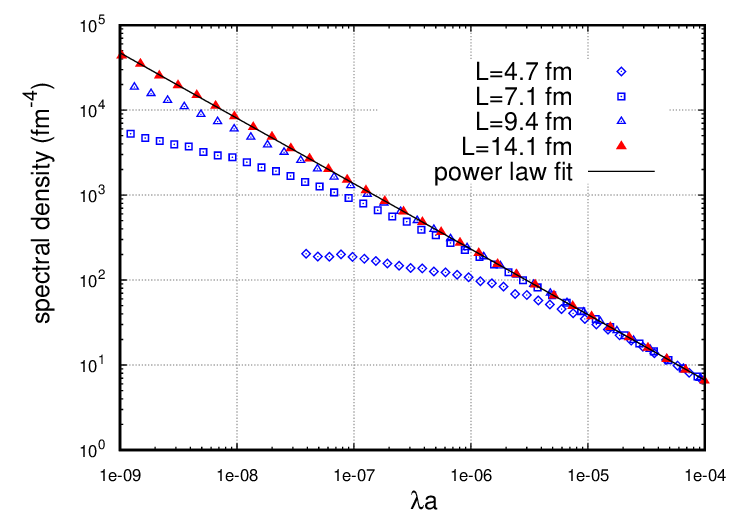} 
  \caption{\label{fig:spd_singular}  The spectral density of the quenched
    matrix model for different system sizes. The continuous line is a
    power-law fit to the common envelope of the curves, corresponding to
    different volumes. }
 \end{centering}
\end{figure}

Using the random matrix model, we can also study the spectral spike in more
detail. As we already saw, independently of the quark mass, the lowest part of
the spectrum can be attributed to a free instanton gas. Therefore, it is
enough to simulate the quenched random matrix model. In the example we show,
the parameters fitted to the quenched $T=1.1T_c$ lattice data were used. In
Fig.~\ref{fig:spd_singular} we show the spectral density of the random matrix
model. The exact zero eigenvalues have been removed from the spectrum, and we
use log-log scale to be able to resolve the smallest eigenvalues. The
different symbols correspond to simulations performed for systems of different
sizes, indicated in the legends. The data points corresponding to different
system sizes possess a common enveloping linear curve. This shows that in the
infinite volume limit the spectral density is described by a singular
power-law. In this particular case a linear fit to the envelope yields
\be
\rho(\lambda) \propto \lambda^\alpha \;\;\; \mbox{with} \;\;\;
  \alpha =-0.770(5),
\ee
which is an integrable singularity.

Our experience shows that if the instanton gas becomes more dilute, the
singularity gets stronger, i.e.\ $\alpha$ moves toward $-1$. We conjecture
that in the chiral limit $\alpha \rightarrow -1$. At first sight this might
seem counter-intuitive, as one would expect light dynamical quarks to repel
the eigenvalues from zero. However, there is no contradiction, since even
though in the chiral limit typical eigenvalues become smaller and the
singularity gets stronger, the total number of eigenvalues in the singular
peak diminishes, and in the chiral limit vanishes, as expected. Singular
behavior of the spectral density at zero in a similar instanton-based model
was already found a long time ago by Sharan and Teper \cite{Sharan:1998vp}.

\section{The fate of chiral symmetry in the chiral limit}

According to the Banks-Casher formula, in the chiral limit the order parameter
of chiral symmetry breaking is the spectral density at zero. We saw that in
the high temperature phase the spectral density is singular at zero, and thus
the Banks-Casher formula cannot be used. However, we can still go back to
Eq.~(\ref{eq:bankscasher}), the derivation of the Banks-Casher
formula. Making use of the fact that even in the chiral limit the eigenvalues
in the singular part of the spectral density are much smaller than the quark
mass, we obtain
\be
\langle \bar{\psi} \psi \rangle \;\; \propto \;\; 
     \langle \sum_i \frac{m}{m^2 + |\lambda_i|^2} \rangle
     \;\; \approx \;\;
       \underbrace{\left( \mbox{\parbox{10ex}{\tiny avg.\ number
             of instantons in free gas}}\right) }_{m^{N\msub{f}} \chi\msub{0} V}
     \times \frac{1}{m} \;\; = \;\; m^{N\msub{f}-1} \chi\msub{0} V.
   \label{eq:pbp}
\ee
Here we used that if $|\lambda_i| \ll m$ then each term in the sum gives a
contribution $1/m$, and after taking the expectation with the path integral,
the average number of terms coincides with the average number of instantons
(plus anti-instantons) in the free instanton gas, responsible for the singular
spike. The upshot is that in the chiral limit the condensate is proportional
to the $N_f-1$-th power of the quark mass. As a result, for more than one
flavor, the condensate vanishes in the chiral limit, as we expect from the
restoration of the flavor non-singlet chiral symmetry. In the case of one
flavor, the condensate does not vanish, but in this case there is no chiral
symmetry to be broken or restored.

In recent years there have been a lot of discussion about how the anomalous
$U(1)_A$ symmetry is manifested in the high temperature phase. A quantity that
was extensively studied is the difference of the pion and delta susceptibility
that is expected to be nonzero if the symmetry is broken. This can also be
written as a Banks-Casher type spectral sum, for which the foregoing argument
can again be used. Thus we obtain that in the chiral limit
\be
\chi_\pi - \chi_\delta \;\;\; = \;\;\;
   \langle \sum_i \frac{m^2}{(m^2 + \lambda_i^2)^2} \rangle
     \;\; \approx \;\; 
       \underbrace{\left( \mbox{\parbox{10ex}{\tiny avg.\ number
             of instantons in free gas}}\right) }_{m^{N\msub{f}} \chi\msub{0} V}
     \cdot \frac{1}{m^2} \; = \; m^{N\msub{f}-2} \chi\msub{0} V.
\ee

This behavior, and that in Eq.~(\ref{eq:pbp}) is fully supported by direct
simulations of our matrix model. Remarkably, for $N_f=2$ the quantity
$\chi_\pi \! - \! \chi_\delta$ is nonzero in the chiral limit. We emphasize that
this happens in spite of the fact that in the chiral limit the topological
susceptibility vanishes, but the anomaly still nontrivially affects the $\pi$
minus $\delta$ susceptibility. This can happen only because the spectral
density is singular at the origin. This scenario has another nontrivial
consequence. Even though the spontaneously broken chiral symmetry is restored
above $T_c$, the order of the thermodynamic and the chiral limit is still
essential. If the chiral limit is taken first in a finite volume, then as can
be seen in Fig.~\ref{fig:spd_singular}, the singularity at the origin is
``regularized'' by the finite volume, and $\chi_\pi - \chi_\delta$ becomes
zero.

This argument also highlights the difficulties in seeing this effect in actual
lattice simulations. The influence of the anomaly on these quantities, related
to chiral symmetry is proportional to the topological susceptibility. Even in
the quenched case the topological susceptibility falls sharply above $T_c$,
and light dynamical quarks suppress topological fluctuations even
more. Therefore, at high temperatures the effect we described can be
numerically small. If the volume is not large enough to contain typically
several instantons and anti-instantons, the spectral spike does not form. In
addition, to resolve the small eigenvalues in the spectral spike, one needs a
chiral Dirac operator both for the sea, to account for the proper suppression
of the small eigenvalues, and for the valence to have the correct contribution
of the small eigenvalues to physical quantities.

\section{Conclusions}

We saw that the singular spike in the spectral density at the origin can be
explained by mixing would be zero modes of a free gas of instantons and
anti-instantons. This is consistent with constraints obtained for the spectral
density from the assumption of the restoration of the spontaneously broken
chiral symmetry \cite{Giordano:2024jnc,Giordano:2024awb}, and also consistent
with the quasi-instanton picture of Kanazawa and Yamamoto
\cite{Kanazawa:2014cua,Kanazawa:2015xna}. As we already remarked, instantons
here do not mean field configurations close to solutions of the field
equations, and they might not even mean well localized lumps of topological
charge. Indeed, all that we can infer from the success of our model is that
the would be zero modes are exponentially localized with exponentially small
mixing matrix elements. Finding different localization properties in the gauge
field and the zero modes would not be unnatural, as in contrast to the zero
mode, the gauge field of the finite temperature caloron solution falls only
like a power-law \cite{Kraan:1998pm}. The topological charge density certainly
has nontrivial structures \cite{Mickley:2023exg}, and exploring their
relationship to the spatial structure of eigenmodes in the ZMZ would be
interesting.

Our instanton-based random matrix model also predicts the appearance of some
tightly bound instanton--anti-instanton pairs, in addition to the free
instanton gas. However, in the chiral limit the splitting of the corresponding
eigenvalues remains constant, governed by the small size of the
molecules. Therefore, the corresponding eigenvalues do not appear in the
singular spike, and they do not contribute to the quantities dominated by the
spectral spike \cite{Kovacs:2023vzi}.

We showed that in one particular quantity, the pion minus delta
susceptibility, for $N_f=2$ light flavors the anomalous chiral symmetry
breaking still shows up at high temperature in the chiral limit. Since our
arguments are valid up to arbitrarily high but finite temperatures, the
anomalous breaking of the $U(1)_A$ symmetry should remain up to arbitrarily
high temperatures. However, the strength of the breaking was seen to be
proportional to the topological susceptibility in the quenched limit at the
given temperature. Since this susceptibility falls sharply with increasing
temperature, the size of the symmetry breaking should also become very small,
but still non-vanishing. This suggests that the $U(1)_A$ symmetry is not
restored, but the manifestation of its breaking for different quantities is
still an open question that has also a bearing on the nature of the transition
in the chiral limit. This is a topic that has been recently extensively
studied both in lattice simulations
\cite{Philipsen:2021qji,Cuteri:2021ikv,Mitra:2024mke,Aarts:2023vsf} and low
energy effective models
\cite{Fejos:2020pcg,Fejos:2022mso,Fejos:2023lvw,Fejos:2024bgl,Braghin:2024lbo}.

\acknowledgments 

The author is supported by NKFIH grant K-147396 and excellence grant
TKP2021-NKTA-64.

\end{document}